\documentclass{emulateapj}

\slugcomment{Accepted for publication in the Astrophysical Journal}

\shorttitle{SCATTERING AND INFLOWING BROAD-LINE REGIONS} \shortauthors{GASKELL AND
GOOSMANN}

\begin{document}

\title{LINE SHIFTS, BROAD-LINE REGION INFLOW, AND THE FEEDING OF AGNS}

\author{C. MARTIN GASKELL\altaffilmark{1,2}}

\altaffiltext{1}{Department of Astronomy, University of Texas, Austin, TX 78712-0259.}

\altaffiltext{2}{Centro de Astrof\'isica de Valpara\'iso y Departamento de F\'isica y Astronom\'ia,  Universidad de Valpara\'iso, Av. Gran Breta\~na 1111, Valpara\'iso, Chile. (Present address) \email{martin.gaskell.astro@gmail.com}}

\author{REN\'E W. GOOSMANN\altaffilmark{3,4,5}}

\altaffiltext{3}{Astronomical Institute of the Academy of Sciences, Bocni II
1401, 14131 Prague, Czech Republic.}

\altaffiltext{4}{Tinsley Visiting Scholar, Department of Astronomy,
University of Texas, Austin, TX 78712-0259.}

\altaffiltext{5}{Observatoire astronomique de Strasbourg, 11 rue de l'Universit\'e, F-67000 Strasbourg, France. (Present address) \email{rene.goosmann@astro.unistra.fr}}

%\email{martin.gaskell.astro@gmail.com}

\begin{abstract}
Velocity-resolved reverberation mapping suggests that the broad-line regions (BLRs) of AGNs can have significant net inflow.  We use the {\it STOKES}  radiative transfer code to show that electron and Rayleigh scattering off the BLR and torus naturally explains the blueshifted profiles of high-ionization lines and the ionization dependence of the blueshifts. This result is insensitive to the geometry of the scattering region.  If correct, this model resolves the long-standing conflict between the absence of outflow implied by velocity-resolved reverberation mapping and the need for outflow if the blueshifting is the result of obscuration.  The accretion rate implied by the inflow is sufficient to power the AGN.  We suggest that the BLR is part of the outer accretion disk and that similar MHD processes are operating.  In the scattering model the blueshifting is proportional to the accretion rate so high-accretion-rate AGNs will show greater high-ionization line blueshifts as is observed.  Scattering can lead to systematically too high black hole mass estimates from the C\,IV line. We note many similarities between narrow-line region (NLR) and BLR blueshiftings, and suggest that NLR blueshiftings have a similar explanation.  Our model explains the higher blueshifts of broad absorption line QSOs if they are more highly inclined.  Rayleigh scattering from the BLR and torus could be more important in the UV than electron scattering for predominantly neutral material around AGNs. The importance of Rayleigh scattering versus electron scattering can be assessed by comparing line profiles at different wavelengths arising from the same emission-line region.
\end{abstract}

%\end{document}


\section{INTRODUCTION}

The structure and kinematics of the broad-line region (BLR) of thermal
active galactic nuclei\footnote{For a review of the differences between thermal and non-thermal
AGNs see \citet{antonucci12}.} (AGNs) has long been a subject of much
debate, and a wide range of structures and velocities have been
considered (for reviews see
\citealt{mathews_capriotti85,osterbrock_mathews86,osterbrock93,gaskell+99,sulentic+araa00, gaskell09}).
Because of these uncertainties, for a long time it was not clear where the BLR
is located, what it is doing, and hence, what role it plays in the
AGN phenomenon.  Reverberation mapping
\citep{lyutyi+cherepashchuk72,cherepashchuk+lyutyi73,blandford_mckee82,gaskell_sparke86} has enabled us to probe
the structures of the BLR and dusty torus, and it is argued elsewhere
(see \citealt{gaskell+07}, hereinafter ``GKN''; and \citealt{gaskell09}) that the
BLR and torus share a similar flattened toroidal structure with a
high covering factor and self shielding.

A major reason for the uncertainty over the structure and kinematics of the BLR has been
the conflicting
pictures of the velocity field in AGNs field given by varying lines of observational evidence.  First, the discovery of broad absorption lines \citep{lynds67} was unequivocal evidence that
at least {\it some} gas is outflowing from AGNs. Radiatively-driven outflows could also
naturally explain the symmetric ``logarithmic'' BLR line profiles seen in a large fraction of AGNs
\citep{blumenthal_mathews75}. The outflow picture was supported by
the discovery \citep{gaskell82} of the blueshifting of the
high-ionization BLR lines with respect to the low-ionization lines
and the rest frame of the host galaxy by $\sim 600$ km s$^{-1}$.
This effect, which for brevity we
will refer to here simply as ``blueshifting'', required physical separation of the high- and
low-ionization BLR clouds, a component of radial motions, and an opacity source. \citet{gaskell82}
proposed a ``disk-wind'' model where the blueshifting could be explained by having the
high-ionization clouds be radially outflowing, with obscuration in
the equatorial plane blocking our view of the receding clouds (i.e.,
of the redshifted side of the line profile).  Although the
blueshifting is usually of the order of $\sim 600$ km s$^{-1}$, it
can exceed 4000 km s$^{-1}$ \citep{corbin90}.  The blueshifting is
not only found when comparing the profiles of high- and
low-ionization lines in individual AGNs, but also when using
spectral principal component analysis (SPCA) of samples of AGNs to
separate out line profiles into possible independent components such as an
``intermediate-line region'' (ILR) and a ``very broad line region''
(VBLR) (see \citealt{brotherton+94}). The magnitude of the
blueshifting is roughly in order of increasing ionization potential
\citep{tytler_fan92}. It also tends to be strongest in luminous,
radio-quiet AGNs \citep{corbin90,tytler_fan92,sulentic+95,richards+02},
especially in broad absorption line QSOs (BALQSOs; \citealt{corbin90,richards+02}),
and in AGNs with a high accretion rate
\citep{sulentic+00,xu+03,leighly_moore04}. The large blueshifts in
high-accretion-rate AGNs have been taken as an indication of strong
outflowing winds in these AGNs (e.g.,
\citealt{leighly_moore04,komossa+08}).

\citet{gaskell88} pointed out that the outflowing high-ionization
BLR scenario predicted strong velocity-dependent time delays in the
wings of the high-ionization lines (the blue wing should lead the
red wing by twice the line--continuum delay), and showed that such a
signature of outflowing winds was absent at a high confidence level
in velocity-resolved reverberation mapping of NGC~4151.  This has
subsequently been found to be the case in many other AGNs where
velocity-dependent time delays have been studied (\citealt{koratkar_gaskell89,crenshaw_blackwell90,koratkar_gaskell91a,koratkar_gaskell91b,korista+95}; see \citealt{gaskell10b} for more detailed discussion).
With only a couple of exceptions discussed below, the wings of BLR lines either vary simultaneously (as would be
expected from Keplerian motion of the clouds or isotropic motions),
or show slight evidence for inflow (i.e., the red wings vary first).
Because of the strong evidence for gravitational domination of the motions,
\citet{gaskell88} argued that motions of BLR clouds
were gravitationally dominated, and hence that they could be used
for determining black hole masses.  Furthermore,
\citet{krolik+91} showed from an analysis of the \citet{clavel+91} observations of NGC~5548 that broad
line widths are consistent with the $r^{-1/2}$ fall off with radius
expected when motions are virialized.  The Krolik et al.\@ result has been confirmed for other objects
\citep{peterson_wandel00,onken_peterson02,kollatschny03} covering a wide range of black hole
masses and Eddington ratios.

There has thus long been little doubt that {\em low}-ionization BLR clouds
(i.e., those producing Mg\,II and the Balmer lines) are
predominantly orbiting the black hole.  Other lines of evidence point to this orbital motion being
Keplerian motion in the equatorial plane.  Despite the significant covering factor of the BLR, we
never see the BLR in absorption (see GKN), Balmer line widths
show the expected correlation with the orientation of the rotation
axis \citep{wills_browne86,rokaki+03}, disk-like line profiles are common
(e.g., \citealt{eracleous_halpern94,gaskell_snedden99}), and evidence of
orbital motion of emission regions has been detected
\citep{gaskell96,sergeev+02,pronik_sergeev06}. It has been suggested that
double-peaked line profiles are due to separate BLRs around two supermassive black holes
in a binary \citep{gaskell83} but variability observations strongly support a disk
origin instead (see \citealt{gaskell10a} and references therein). \citet{gaskell10c,gaskell11}
shows that apparently extreme double BLR peaks can readily by explained with the same gas
distribution and kinematics as for single-peaked BLRs seen from a slightly higher inclination.
The question of the detectability of close supermassive binary black holes remains an active area of investigation, however (see \citealt{popovic12} for an extensive review).

It can be seen that there is considerable confidence that black hole masses can reliably be
estimated from {\em low}-ionization lines (see \citealt{marziani+sulentic12} for a review
of AGN black hole mass determinations).  However, for the {\em
high}-ionization lines, the conflict between the kinematics implied
by the blueshifted absorption lines and the blueshifting of the
emission lines on the one hand, and the velocity-resolved
reverberation mapping on the other, has raised serious doubts about the
suitability of high-ionization lines such as C\,IV $\lambda$1549 for
estimating black hole masses.  Further evidence for a difference in
the kinematics of high- and low-ionization lines comes from the
different velocity dependencies of the physical conditions
\citep{snedden_gaskell04}.  The ionizing flux
received by low-ionization BLR lines shows the dependence on
velocity one would expect as a result of virialization and the
inverse-square law, while the ionizing flux received by the
high-ionization clouds appears to be almost independent of velocity.

This conflict between BLR kinematic indicators cannot simply be reconciled
by assuming that the low-ionization lines arise in a disk while the
high-ionization lines arise in a wind (e.g.,
\citealt{collin-souffrin+88}) because the velocity-resolved
reverberation mapping specifically shows
\citep{gaskell88,koratkar_gaskell89,crenshaw_blackwell90,koratkar_gaskell91a}
that the {\em high}-ionization
BLR gas is {\it not} outflowing. Also, the $r^{-1/2}$ fall off of line widths with radius
\citep{krolik+91} includes the high-ionization C\,IV line.  Furthermore, disk--wind models have a
problem of explaining why line profiles of different ions are so
similar if they have very different origins \citep{tytler_fan92}.
While winds necessarily exist in AGNs in order to remove angular momentum
so that material can accrete, and they could be energetically
important as well, the amount of mass involved is small.  The density in a wind is
more than an order of magnitude lower than in the disk, and emissivity
goes as the square of the density, so emission from a wind is negligible.

In this paper we will argue that the velocity-resolved reverberation
mapping results are correct, and that the {\em entire} BLR has a net
inflow. In section 2 we summarize the evidence for an inward spiraling of
the BLR and estimate the inflow component of velocity.  In section 3 we argue
that scattering is consistent with producing a blueshifting
when the BLR is inflowing, and in section 4 we use the {\it STOKES}
Monte Carlo radiative transfer code to show that scattering off an
inflowing medium reproduces observed blueshifted line profiles and
also explains the dependence of blueshifting on ionization. We
consider the implications of the inflowing BLR scenario for the
energy generation mechanism in AGNs in section 5.  In section 6 we
offer an explanation of why high-accretion-rate AGNs (``Narrow-line
Seyfert 1s'' = NLS1s) show a stronger blueshifting, we point out
potential systematic effects when the C\,IV emission line is used to
estimate masses of high-redshift AGNs, we suggest an extension of our
model to the narrow-line region, and we offer an explanation of
why the blueshifting is enhanced in broad absorption line QSOs.

\section{Inflowing BLR Gas}

\subsection{The Evidence for an Inflow Velocity Component}

Although it has generally been assumed in estimating black hole
masses that the BLR gas is in near-Keplerian or quasi-random orbits (i.e., there is no net radial motion),
\citet{gaskell88} found that for NGC~4151 {\em inflow} was favored and that
purely Keplerian or random orbits were excluded at the 97\% (single-tailed)
confidence level. \citet{koratkar_gaskell89} similarly found that
non-inflowing motion was excluded in Fairall 9 at the $\sim 95$\%
confidence level. From the intensive 1989 {\it IUE}
monitoring of NGC~5548 \citep{clavel+91},
\citet{crenshaw_blackwell90} and \citet{done_krolik96} also favored
a net inflow of the C\,IV emitting gas in NGC~5548.  From the estimated
errors in the lags given by \citet{crenshaw_blackwell90},
non-inflowing motion is excluded at the 93\% confidence level. The
1993 combined {\it HST} and {\it IUE} campaign \citep{korista+95}
showed a similar degree of inflow (see their Table 25).  The only bright AGN for which the
variability of C\,IV has been studied is 3C~273 where \citet{koratkar_gaskell91b} and \citet{paltani+turler03} find an
inflowing velocity component at two separate epochs. As
\citet{gaskell_snedden97} point out, although the statistical
significance of inflow C\,IV for one line in any one observing
campaign of any individual AGN is not necessarily strong, the case
is much stronger when all the AGNs are considered
together.\footnote{\citet{koratkar_gaskell91b} also find
non-statistically significant C\,IV inflow in four additional AGNs.}
The aforementioned analyses are mostly for C\,IV $\lambda$ 1549, but
\citet{gaskell88} found inflow of Mg\,II $\lambda$2798 in NGC~4151,
and \citet{welsh+07} have recently found that H$\beta$ in NGC~5548
has an inflow component of velocity.

Two contradictory velocity-resolved reverberation mapping results must be mentioned.
The first is an event in NGC~5548 which {\em temporarily} showed an apparent {\em outflow} kinematic
signature \citep{kollatschny+dietrich96}.  The second is a similar conflicting signature seen in one season of
monitoring of NGC~3227 \citep{denney+09}.  The temporary NGC~5548 apparent outflow signature certainly does {\em not}
represent a systematic outflow because only three months later an {\em inflow} signature was seen in the same object \citep{kollatschny+dietrich96}.   \citet{gaskell10c,gaskell11} has shown that transient
apparent outflow signatures are a natural consequence of off-axis variability for which there is considerable other evidence.

Independent evidence for inflow comes from high-resolution
spectropolarimetry \citep{smith+05}.  The systematic change in
polarization as a function of velocity across the Balmer lines
suggests a net inflow of a scattering region somewhat exterior to
the Balmer lines.

In summary, we believe that there is significant observational evidence for gas producing
both the high- and low-ionization BLR lines to be inflowing.

\subsection{The Inflow Velocity}

We can estimate the ratio of net inflow velocities to random
velocities along the line of sight from the positions of the peaks
of the red-wing/blue-wing cross correlation functions (CCFs).  The
positions of the expected peaks in the wing-wing CCFs are marked in
\citet{gaskell88} and \citet{koratkar_gaskell89}.  The observed peak
positions, and the one found by \citet{crenshaw_blackwell90} are all
consistent with the net inflow velocities being several times
smaller than the non-inflow velocities. \citet{done_krolik96} reach
a similar conclusion from more detailed modelling of the
velocity-dependent delays in NGC~5548. For each object this suggests
that the net inflow velocity of C\,IV is $\sim 1000$ km s$^{-1}$. In
a totally independent analysis of the polarization structure of
Balmer lines, \citet{smith+05} suggest an inflow velocity of the
scattering region of 900 km s$^{-1}$. We will therefore adopt an
inflow velocity of 1000 km s$^{-1}$.  However, we will see below that the
precise value of the inflow velocity is not critical for our modelling and that
the blueshifting can be obtained with much lower velocities.

\section{Producing a Blueshift from Inflowing Gas}

There are two main ways of producing a blueshift of a line from
inflowing gas.  One is by having anisotropic emission from the gas
clouds \citep{gaskell82,wilkes84}, and the other is by having
scattering off inflowing material (e.g.,
\citealt{auer_vanblerkom72}).  Although the emission from BLR clouds
is certainly expected to be anisotropic, the anisotropy is much
greater for some lines than for others. As discussed by
\citet{wilkes_carswell82} and \citet{kallman+93}, this creates a
problem in explaining {\it all} line profiles with an inflowing
anisotropic emitting cloud model. Lyman $\alpha$ has particularly
strongly asymmetric emission compared with other lines when clouds
are optically thick, which is almost certainly the case for the BLR
clouds of relevance here (see \citealt{snedden_gaskell07} for
evidence against a significant optically-thin contribution to the
BLR). There is no evidence that Lyman $\alpha$ is more asymmetric
than other lines \citep{wilkes_carswell82}, so it
would be hard for anisotropic emission from inflowing clouds to
explain the blueshifting.

Electron scattering has long been considered to be a significant
source of line broadening in AGNs in general
\citep{kaneko_ohtani68,weymann70,mathis70}, and from time to time it
has been invoked to explain the line profiles of individual objects
\citep{shields_mckee81,laor06}.\footnote{Note however, that the
apparent extended wings in the AGN considered by
\citet{shields_mckee81} are actually due to red continuum being too
low in the \citet{baldwin75} spectrum (J. M. Shuder - private
communication).} Scattering regions with a net radial motion were
considered by \citet{auer_vanblerkom72}. If the scattering region is
outflowing, scattered photons are redshifted, while if it is
inflowing, the photons are blueshifted.  This can easily be understood by considering
one´s reflection in a moving mirror (see Fig.~10 of \citealt{gaskell09}).  If the mirror is moving
towards you your image appears to be approaching at twice
the speed of the mirror. Such shifts have already
been shown in simulations of electron scattering in AGNs by
\citet{kallman_krolik86} and \citet{ferrara_pietrini93} and it
has been suggested that this could be a cause of the blueshifting of
high-ionization lines \citep{corbin90,mathews93}.

\section{Scattering in an Inflowing Medium}

\subsection{Spherical Scattering Shells}

We have modelled the effects of an inflowing scattering medium using
the {\it STOKES} Monte Carlo radiative transfer code, which is
described in \citet{goosmann_gaskell07} and \citet{marin+12}. Detailed documentation,
sample input, source code, and executables for different computer
platforms can be freely downloaded.\footnote{{\it
www.stokes-program.info}} The optical depths, $\tau_{es}$, to
scattering by free electrons along our line of sight are not expected to be much greater
than unity inside the high-ionization BLR of an AGN.
\citet{shields_mckee81} estimate $\tau_{es} \lesssim 1$ for
electrons between BLR clouds, and \citet{laor06} estimates
$\tau_{es} \thickapprox 0.3$ for typical BLR clouds. Modelling of
BLR clouds with the photoionization code {\it CLOUDY}
\citep{ferland+98} gave similar values of $\tau_{es}$. We therefore
investigated quasi-spherical external scattering regions with $\tau$
= 0.5, 1, and 2.  We also modelled $\tau$ = 10 to investigate
effects of significantly larger optical depths. The inflow velocity
was taken to be 1000 km s$^{-1}$ in all models. Since in this paper
we are interested in modeling just the inflow and line asymmetry, we
simply assumed that the unscattered line had an intrinsic broadening
due to bulk motions.  The assumed unbroadened profile is shown as a solid black
line in Fig.~1.  We considered inflowing scattering shells both
inside and outside the BLR.  The results are shown in Fig.\@ 1.

% Fig.\@ 1
\begin{figure}
\vspace*{0.3cm}
\includegraphics[width=84mm]{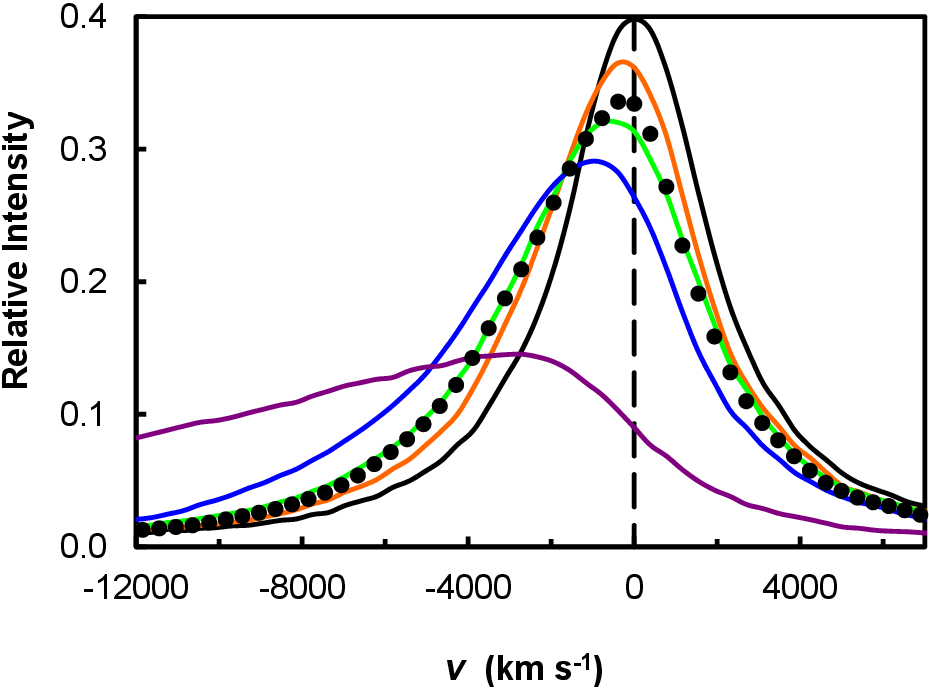} %\vspace*{0.3cm}
\caption{Calculated shifts of the C\,IV line by electron or Rayleigh
scattering. The top black solid curve shows a Lorentzian line
profile before scattering. The other solid curves show the
blueshifting caused by an external spherical shell of scatterers
with an inflow velocity of 1000 km s$^{-1}$. The areas under all
curves are the same.  In order of increasing blueshifting and
decreasing peak flux, the curves show the effects of $\tau$ = 0.5
(red), 1.0 (green), 2.0 (blue), and 10 (purple). The dots show the
blueshifting caused by an inflowing flared disk with a radial optical depth of 5, a half-opening angle of 60 degrees,
and an inflow velocity of 1000 km s$^{-1}$,
when the BLR is viewed from a ``type-1'' viewing position.} \label{electron
scatterings}
\end{figure}
% This figure comes from line_flux_profiles.xls sheet "Plotting 2012"

Our first result is that for a line-emitting region {\em outside}
an inflowing scattering shell there was a negligible effect on the
line profile (differences mostly smaller than the plotting lines and
symbols in the figures).  This thus verifies that when there is
strong radial stratification of the BLR (see GKN and \citealt{gaskell09}), only the
innermost high-ionization lines (those within the scattering region)
are blueshifted.  This gives a natural
explanation of the difference in blueshift with ionization and is a big
advantage of our scattering model over the
outflowing-wind-plus-obscuration model of \citet{gaskell82} because
in the latter model one has to contrive to have the obscuration
affect the outer low-ionization lines less. In our scattering model
the blueshift of a line only depends on the optical depth and
velocity of material {\em outside} where the line is emitted.  As
discussed in section 4.4 below, we can therefore easily predict
blueshifting as a function of ionization.  For example, C\,III]
$\lambda$1909 will have a blueshifting about half that of C\,IV
$\lambda$1549, as is observed to be the case
\citep{corbin90,steidel_sargent91}.

In Fig.\@ 2 we show a comparison with the blueshifted C\,IV profiles in a typical
AGN, and in Fig.\@ 3 we show a comparison with the VBLR line profile given
by SPCA of a sample of AGNs
by \citet{brotherton+94} in.  In Fig.\@ 4
we show the blueshifts for composite SDSS spectra \citep{richards+02} sorted by blueshift.

It can be seen in Fig.\@ 1 that for a spherical shell, the blueshifting and asymmetry
increase with $\tau$.  For $\tau \sim 10$ (a much greater
electron-scattering optical depth than has been considered for the line of sight to a
BLR), both the blueshifting and the asymmetry become much greater
than is observed (see Figs.\@ 2 -- 4), so we can rule out such high optical depths along
the line of sight.

% Fig.\@ 2
\begin{figure}
\vspace*{0.3cm}
\includegraphics[width=84mm]{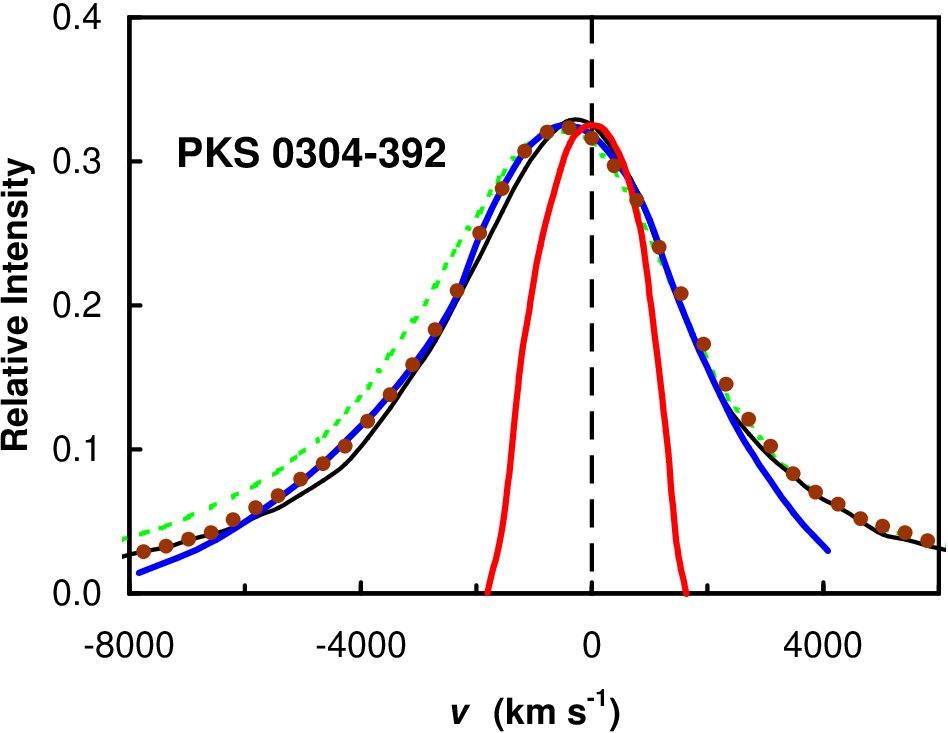} %\vspace*{0.3cm}
\caption{The profiles of O I $\lambda$1305 (narrow symmetric profile
shown in red) and C\,IV $\lambda$1549 (thick blue line) for the
quasar PKS~0304-392.  The thin black line is the blueshifted profile
produced by a spherical distribution of scatters with $\tau$ = 0.5,
and the dashed green line is the profile produced by the same
distribution with $\tau$ = 1.  The brown dots are the profile
produced by a $\tau$ = 20 inflowing cylindrical distribution.
PKS~0304-392 observations taken from \citet{wilkes84}.} \label{PKS_fit}
\end{figure}
% This figure comes from line_flux_profiles.xls sheet "Plotting 2011"

% Fig.\@ 3
\begin{figure}
\vspace*{0.3cm}
\includegraphics[width=84mm]{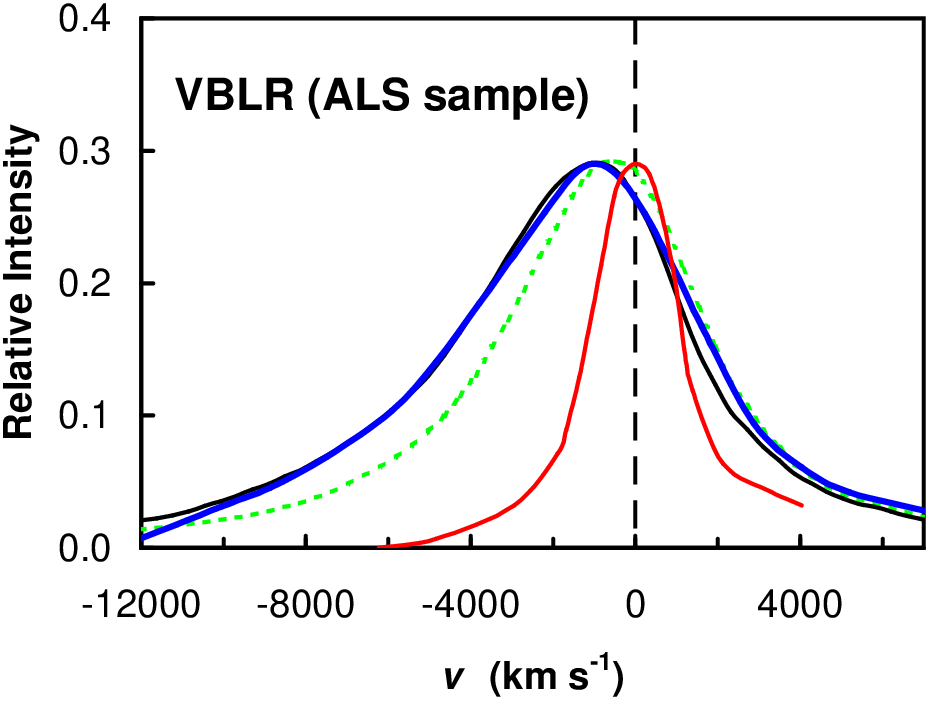} %\vspace*{0.3cm}
\caption{The mean intermediate line region (ILR) profiles (narrow
red profile), and very broad line region (VBLR) profiles (thick blue
line) given by SPCA of the ALS sample of \citet{brotherton+94}.  The solid line
shows the shifting caused by an external spherical shell of scatters
with $\tau$ = 2 inflowing at 1000 km s$^{-1}$ as shown in Fig.\@ 1.
The dotted green line is for a similar shell with $\tau$ = 1.}
\label{vblr_fit}
\end{figure}

% Fig.\@ 4
\begin{figure}
\vspace*{0.3cm}
\includegraphics[width=84mm]{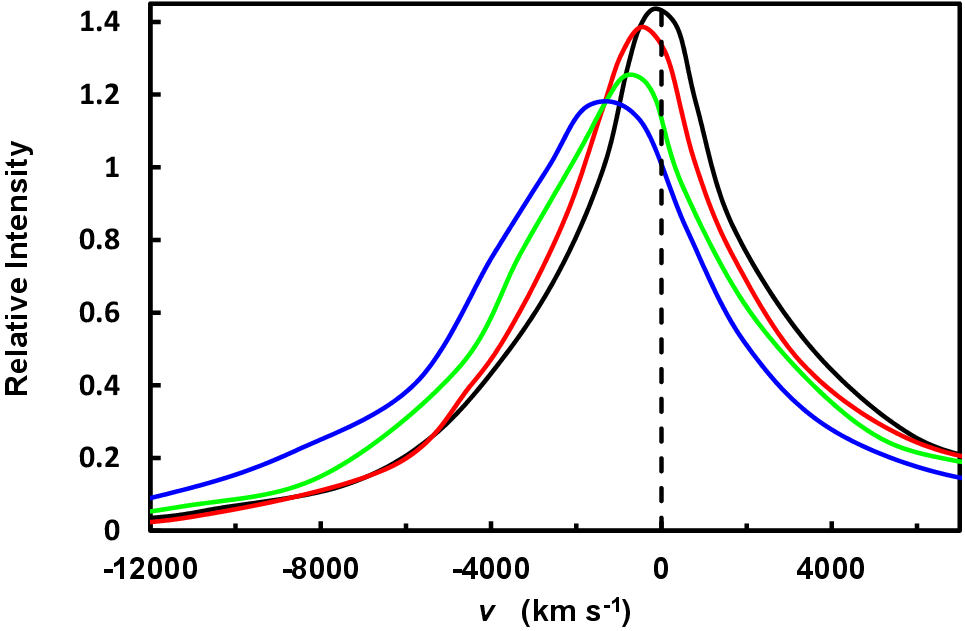} %\vspace*{0.3cm}
\caption{Shifts in composite SDSS spectra binned by C\,IV blueshift.  Each curve is based on
approximately 50 AGNs (see \citealt{richards+02} for details).  The blueshifts of each sample
are 197, 606, 1003, and 1526 km s$^{-1}$ for the black, red, green, and blue curves respectively.
No attempt has been made to remove blending with He\,II $\lambda$1640 or O\,III] $\lambda$1663.
To match Fig\@ 1, the areas under all curves are the same -- i.e., they have been normalized to
the same combined flux in the C\,IV, He\,II, and O\,III] blend.  These line profiles can be compared
with the model predictions of Fig.\,1.}
\label{SDSS_fit}
\end{figure}

\subsection{Rayleigh Scattering in the BLR and Torus}

While large electron scattering optical depths are not likely along our line of sight to the
BLR or within the
inner BLR, our modelling of BLR clouds with {\it CLOUDY} showed that
for clouds of sufficiently large neutral column densities to produce
the significant Fe\,II emission seen in many AGNs, Rayleigh scattering was over an order
of magnitude more important than electron scattering in the UV. This
is not surprising for a region where hydrogen, which is the main
provider of electrons, is mostly neutral, since a typical Rayleigh
cross section is larger in the UV than the Thomson cross section.
The potential importance of Rayleigh scattering in modifying the
spectra of AGNs has been discussed in detail by
\citet{korista_ferland98}, and \citet{lee05} has considered the
effect of Rayleigh scattering on the polarization of Lyman $\alpha$.

{\it STOKES} does not explicitly handle Rayleigh scattering at
present, but the angular dependence of Rayleigh scattering and
electron scattering is identical, and the variation of scattering
cross section across an emission line is unimportant, so Rayleigh
scattering can be treated as electron scattering.  The effect of
Rayleigh scattering off an inflowing medium completely covering the
source will be to give a very large blueshift as shown in Fig.\@ 1 for
$\tau \sim 10$.  Such Rayleigh scattering would, however, be
accompanied by substantial blueshifted low-ionization absorption
which is never observed.  We can therefore strongly rule out such a
{\em spherical} shell which is optically thick to Rayleigh
scattering.

\subsection{Scattering off a Flattened BLR and Torus}

Although it is unlikely that our direct line of sight to the BLR has
a high optical depth to electron scattering, as one goes away from
the black hole the BLR must become optically thick (in order to explain
the strong optical Fe\,II emission) and merge with
the torus (see GKN), so the optical depth to both electron and Rayleigh
scattering will become substantial in the equatorial plane
\citep{korista_ferland98}. The existence of Compton-reflection humps
\citep{pounds+90} is evidence that that there are substantial
electron scattering optical depths. Reprocessing of X-rays requires
$\tau_{es} > 1$ (see \citealt{goosmann+07}). It is also well known
that half of all Seyfert 2 galaxies are observed to be ``Compton
thick'' \citep{risaliti+99}, so most or all AGNs are probably
optically thick to electron scattering in the equatorial plane.
\citet{gaskell+07} have argued that in a typical AGN, the BLR, like
the torus, has a covering factor of 40\% or so. As
\citet{smith+04} and \citet{smith+05} point out, a significant scale
height of the scattering region is also necessary in order to
explain the polarization of type-1 AGNs.  The broad-band polarization properties
as a function of opening angle are considered in detail in \citet{marin+12}.
We therefore modelled an inflowing scattering cylindrical torus with a half-opening angle of
60 degrees, an optical depth of 5, and an inflow velocity of 1000
km s$^{-1}$. The viewing position is within the half-opening angle (i.e., as required
for a type-1 viewing position).
We show the profiles arising from such a model by the dotted curve
in Figs. 1.

It can be seen from Fig.\@ 1 that the blueshifting produced by such a model
differs insignificantly from that produced by the purely spherical
scattering model with $\tau \sim 1$ or 2.  However, unlike the spherical
model, the shift produced in the torus model depends only on the
inflow velocity, and except for small opening angles (where the flared disk begins to
approximate a shell) the shift is less sensitive to large optical depths.
This is because in the torus case, after one or two scatterings a
photon has a high probability of escape within the half-opening
angle.  The main difference between the quasi-spherical scattering
case and the torus case is that the former produces more blueward
asymmetry for a given inflow velocity and optical depths.  For most objects, such as
the AGN shown in Fig.\@ 2, the difference between the two models is negligible.  The advantages of
such a flattened BLR and torus are that there is no need to fine
tune $\tau$, and high Rayleigh scattering optical depths are
permitted.

The precise geometry of the inflowing region is not important, only the covering factor.  As an
illustration of this the dots in Fig.\@ 2 show the profile resulting from scattering off an optically-thick
cylindrical inflow.  This can again be seen to give a line profile similar to an inflowing shell of $\tau \sim$ 1.
The similarity of line profiles for a range of geometries is easy to understand because the observed profile is
simply the sum of the unscattered profile plus a scattered blueshifted profile.  The size of the scattered
component depends on the covering factor, and its shift depends on the inflow velocity and the number of
scatterings.

Note that in all our models there is some degeneracy between inflow velocity and covering factor, and,
if the covering factor is large, with the optical
depth as well (see Figs.~1 and 2).  This means that our adopted $v_{inflow}$ of 1000
km s$^{-1}$ is not critical.  If $v_{inflow} < 1000$ km s$^{-1}$ the observed blueshifts can
readily be reproduced by increasing the covering factor and/or $\tau$ of the scattering medium.  For example,
with a scattering region with a half-opening angle of 60 degrees $v_{inflow}$ = 700 km s$^{-1}$
and $\tau_{scatt} = 1$ gives essentially identical profiles as  $v_{inflow}$ = 100 km s$^{-1}$
and $\tau_{scatt} = 10$.  On the other hand, it is not likely that $v_{inflow} > 1000$ km s$^{-1}$ or
else the blueshifting predicted will exceed the observed shifts.

\subsection{Blueshifting as a Function of Emission Radius}

Our scattering model predicts that the blueshift of a line depends
on the inflow velocity immediately outside the radius, $r$, the line
is produced at. The radii that given lines are produced at are known
approximately from reverberation mapping, and the GKN self-shielding
photoionization model also predicts relative radii which agree well
with the observed radii for NGC~5548 and other objects.  For an inflowing BLR we
expect the net inflow velocity, $v \propto r^{-0.5}$.  In Fig.\@ 5 we
compare the relative radii, $R$, at which lines are emitted in the GKN
model with the mean observed blueshifting velocities for AGNs. The
mean blueshift velocity of C\,III] $\lambda$1909 has been taken as
half the blueshifting of C\,IV relative to Mg II (see Fig.\@ 3 of
\citealt{corbin90}), and other blueshift velocities have been taken
from \citet{tytler_fan92}.  The predicted radii (in light days) are
scaled to NGC~5548.  While the blueshifts are for the large samples
of AGNs studied by \citet{corbin90} and \citet{tytler_fan92} rather
than for NGC~5548 itself, there is a similar correlation if the mean
blueshifts are plotted against the measured lags for NGC~5548. The
least-squares fit line in Fig.\@ 5 gives $v \propto R^{-0.52}$ which
is consistent with the expected slope of -0.50.

% Fig.\@ 5
\begin{figure}
\vspace*{0.3cm}
\includegraphics[width=84mm]{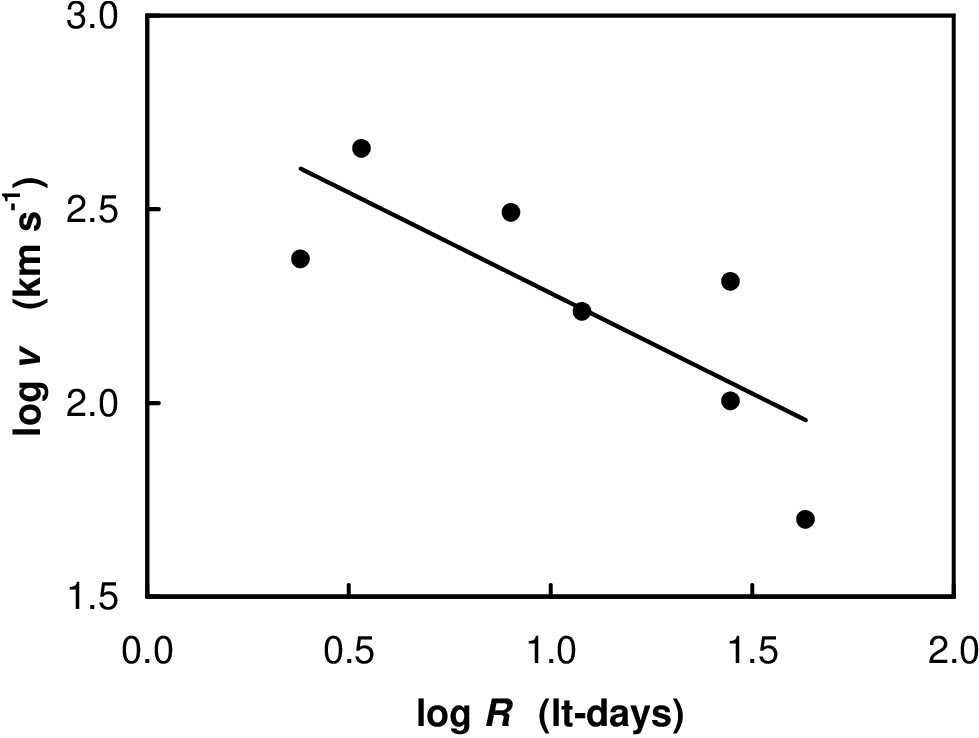} %\vspace*{0.3cm}
\caption{Mean blueshifts of emission lines (see text for details)
versus the radii, $R$, of maximum emission predicted in the GKN
model. The line is a least squares fit showing $\log v$ as a
function of $\log R$.} \label{relative_shifts}
\end{figure}

The SDSS sample of \citet{richards+02} provides additional support for
increasing blueshifting with ionization.   Fig.~6 shows the He\,II $\lambda$1640 and C\,IV $\lambda 1549$ blueshifts
which we have derived from the composite SDSS profiles.  It can be see that for the sample with the highest blueshifts, the He\,II $\lambda$1640 blueshift is approximately twice that of the C\,IV $\lambda 1549$ line as is predicted from the GKN model.

% Fig.\@ 6
\begin{figure}
\vspace*{0.3cm}
\includegraphics[width=84mm]{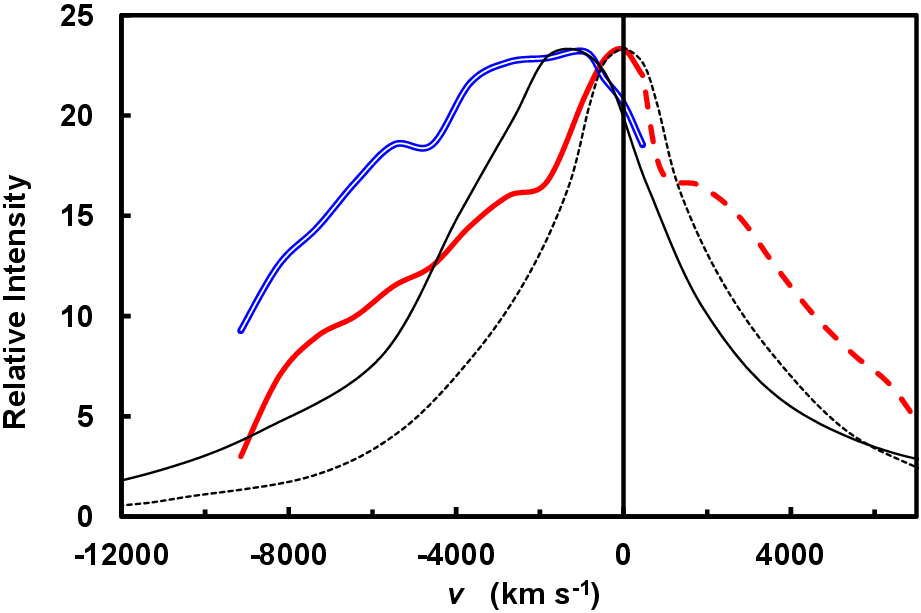} %\vspace*{0.3cm}
\caption{He\,II $\lambda$1640 profiles estimated from composite profiles of SDSS AGNs with
low blueshifting (red lines) and high blueshifting (double blue line).  The C\,IV profiles for
the same samples are shown as dotted and solid thin black lines respectively.   The red side of He\,II $\lambda$1640
is heavily contaminated by O\,III] $\lambda$1640 and is shown by the dashed red line.  It is similar for both samples.  The high-velocity blue wing of He\,II is dominated by C\,IV.}
\label{He II}
\end{figure}

\section{IMPLICATIONS}

\subsection{Mass Inflow Rate}

It has long been noted that, for pure radial motion,
the mass transfer rate across the BLR is comparable to the accretion
rate. \citet{padovani_rafanelli88} have argued, for example, that if
the BLR is purely inflowing, then the mass inflow rate in AGNs
correlates well with the accretion rate. Calculating the mass inflow
rate is straight forward.  We illustrate this by considering the
C\,IV emitting region of the well-studied AGN NGC~5548. The sizes of
the emitting regions are known in NGC~5548 from reverberation
mapping and the continuum shape is also relatively well known. After
reddening correction \citep{gaskell+04,gaskell_benker07}, GKN get a
bolometric luminosity of $10^{45.07}$ ergs s$^{-1}$ during the high
state of NGC~5548 observed by \citet{korista+95}. (This is only
slightly greater than the $10^{44.85}$ ergs s$^{-1}$ estimated by
\citealt{padovani_rafanelli88}). Using the GKN continuum, an
observed C\,IV-emitting radius of 8 light-days, and an electron
density $n_H = 10^{10}$, {\it CLOUDY} photoionization models give a
thickness of $10^{13.7}$ cm for the C\,IV emitting region, and a
mass of 1 solar mass.  For an inflow velocity of 1000 km
s$^{-1}$ the inflow time from 8 light days is $\sim 6$ years.
Adopting a 50\% covering factor (see GKN), this gives a mass inflow
rate of 0.08 solar masses per year.

If we adopt a black hole mass of $10^{7.9}$ solar masses from the
average of the many NGC~5548 black hole mass estimates given by
\citet{vestergaard_peterson06}, and assume a standard radiative
efficiency of 10\%, then the Eddington accretion rate is 0.8 solar
masses per year, and the accretion rate during the high state
studied by GKN is 0.1 solar masses per year.  This is comparable to
the accretion rate we have calculated from the C\,IV emitting gas,
so we can see that the BLR mass inflow can readily provide the mass
inflow rate needed to power NGC~5548.
Since the size of the C\,IV emitting region is well known from
reverberation mapping and we know the covering factor to $\pm 50$\%,
the main uncertainty in this
calculation is the mean density.  If the inflow velocities is less than
hundreds of km s$^{-1}$ then one requires mean densities of the inflowing
material of $10^{12}$~cm$^{-3}$ or more.

Although we have calculated the mass inflow rate just for NGC~5548,
it should be noted that there is no reason to think that NGC~5548 is
unusual in this regard. \citet{padovani_rafanelli88} have shown that
their estimated mass flow rates (calculated assuming that the entire
BLR is moving radially) are proportional to the accretion
rate needed to produce the bolometric luminosity for a wide variety
of AGNs, including objects such as I Zw 1, which we now recognize as
a high-accretion-rate narrow-line Seyfert 1.

\subsection{Producing Inflow}

Because of the degeneracy with the optical depth the inflow velocity is not well
constrained.  However, the inflow velocity needed to explain the blueshifting is likely to be
higher than the sound
speed at the radius of the BLR.  Is it possible to obtain such a net inflow velocity?
In the classic Bondi solution there is no problem in having supersonic infall at large distances,
but this is unlikely to be relevant to what we are considering since the
accreted material has angular momentum.

As is well known, to get a net inflow of matter there must be
outward transport of angular momentum.  This means that something must apply a
torque on the gas.  This could be magnetic breaking caused by a wind \citep{Blandford+Payne82}
or a viscosity providing a torque between material at different radii. The nature of the
viscosity was a long-standing
problem for accretion disk modelling, since ordinary gas
viscosity is far too low, but it is now recognized that the
necessary viscosity comes from the magneto-rotational instability
(MRI) \citep{balbus_hawley91}.

If we have inflow in the BLR, there is the same need for a torque to provide the
necessary angular momentum transfer in the BLR gas.  It would be natural for this to provided by
the same mechanisms as for the disk, i.e., magnetic breaking by a wind, or the MRI.

Visual inspection of MHD simulations of AGN accretion suggests
that the turbulent and inflow velocities are roughly of the same magnitude,
which is in qualitative agreement with our estimate of the inflow velocity.  However,
while there can be highly supersonic inflow in the ``plunging region'' (a few
Schwarzschild radii from the black hole), the turbulent and inflow velocities in
typical MHD simulations of rotationally-supported Keplerian disks are subsonic at the radius
of the BLR.  There are however some simulations of gas with sub-Keplerian rotation with substantial
inflow velocities.  One example is \citet{Proga+Begelman03} where the inflow velocity in the outer
part of the simulations is $\sim$ 10\% of the Keplerian velocity (see the upper right panels in
their figures 9 and 10).  Another possibility \citep{Giri+Chakrabarti13} is that there is a two-component
advective flow where a Keplerian disk is surrounded by a rapidly infalling sub-Keplerian halo.

Flares due to the tidal disruption of stars are a valuable laboratory for testing models of
accretion disks because they arise from a well-defined injection of mass ($\sim$ one solar mass) in
a one-off event. \citet{Cannizzo+90} used a time-dependent alpha-disk model to study the
subsequent evolution of the accretion disk created.  They concluded that the disk remained
luminous for several thousand years.  However, more recent work by \citet{Montesinos+11} shows
that the duration of the flare (and disk) is of the order of only a few years to a few decades.
This much shorter lifetime means that the inflow velocity is a couple of orders of magnitude higher than
in the \citet{Cannizzo+90} model.  Now that a number of tidal disruption events have been observed
we know that the duration is as in the \citet{Montesinos+11} models and hence that the inflow velocities
are higher than in earlier models.

Although there is theoretical and observational support for relatively rapid (supersonic) inflow,
more modelling of the motions around supermassive black holes is clearly needed.

\subsection{The Relationship Between the BLR and the Accretion Disk}

It is now recognized that we are viewing most type-1 AGNs within
$\sim 45$ degrees of the axis (see \citealt{antonucci93} for a
review), and hence that we view the BLR close to face on.  Even when
we see disk-like line profiles, the inclinations are still not great
(see, for example, \citealt{eracleous_halpern94} and Fig.\@ 2 of \citealt{gaskell11}).
Because the observed broad-line widths are substantial, and the BLR has to be
flattened (see GKN), there {\it must} be a {\em substantial}
component of random BLR velocity out of the plane \citep{gaskell09}.  Such a velocity
is also necessary to maintain the thickness of the BLR needed to
provide the observed covering factor.

The overall motion of the BLR (see \citealt{gaskell09}) has to be as follows: the
main motion is rotational but there is a vertical random
(``turbulent'') component of velocity \citep{osterbrock78} that is
almost as great as the rotational velocity. Then, as argued above,
there is an additional, slower, net inflow. As we have noted, this overall structure
of the velocity field in the BLR is qualitatively similar to the
results of accretion disk magnetohydrodynamic (MHD) calculations (see, for example
the simulations of \citealt{hawley_krolik01}).  \citet{gaskell08} points out
that the size of the accretion disk is such that its outer
parts (those generating the optical and IR emission) must extend out
to within the BLR. We therefore propose on the basis of the
similarities of the size, physical processes needed, kinematics, and
mass inflow rates, that {\em the BLR and the outer part of the
accretion disk are one and the same}. The accretion disk
is the part which is optically thick in the continuum and
closest to the equatorial plane.

The magnetic fields generated by the MRI potentially solve two major
problems.  The first is the long-standing ``confinement problem''
(see \citealt{mathews_capriotti85} for a review).  \citet{rees87}
has shown how magnetic fields can confine the BLR clouds. A second
problem is the survival problem.  Clouds with a random velocity
component (e.g., as proposed by \citealt{osterbrock78}) have a mean
time between collisions comparable to the orbital timescale (see
\citealt{osterbrock_mathews86}).  Cloud--cloud collisions will
produce very high Mach number shocks which will immediately destroy
the colliding clouds. Strong magnetic fields can prevent shocks from
occurring, just as they prevent such shocks in MHD simulations
of accretion disks.

\section{DISCUSSION}

\subsection{High-Accretion-Rate AGNs}

As discussed in the introduction, it is well established that
high-accretion-rate AGNs (such as NLS1s) show strong blueshifting,
and this has been interpreted as evidence for strong outflowing
winds in high-accretion-rate AGNs (e.g.,
\citealt{leighly_moore04,komossa+08}). However, the blueshiftings in
NLS1s are merely the extreme of the distribution for AGNs in
general, so there is no reason to think that they have a different
cause from the shifts in other AGNs. We therefore propose that
the greater blueshifts in NLS1 imply greater {\em inflow} rates.
This result follows quite naturally from our {\it STOKES} modelling
shown in Fig.~1. The amplitude of the blueshifting obviously depends
linearly on the inflow velocity, but as shown in Fig.\@ 1, it also
depends on the optical depth and covering factor.  An inflow velocity
of 1000 km s$^{-1}$, as we
have assumed above, is adequate to explain the typical
blueshiftings, but for the most extreme examples, such as Q1338+416,
where the shift is almost 5000 km s$^{-1}$ \citep{corbin90}, a
higher inflow velocity is needed.  In Fig.~1 the blueshifting is
roughly proportional to the product of the scattering optical depth
and the inflow velocity. For a given column length the optical depth
depends on the filling factor and density. The mass inflow rate is
proportional to the mean density and inflow rate. Thus, whether a greater
optical depth or a greater inflow velocity is responsible for the
increased blueshifts, {\em the blueshift is proportional to the mass
inflow rate}.

\subsection{Estimating Black Hole Masses from C\,IV Widths}

There is considerable interest in estimating black holes masses at
high redshifts.  Because it is difficult to measure H$\beta$ at high
redshift, it has been suggested that the FWHM of C\,IV $\lambda$1549
can be used instead of the FWHM of H$\beta$
\citep{vestergaard02,warner+03}.  Our conclusion that the
C\,IV-producing region has a net inflow rather than an outflow in a wind
is good news for such endeavors.  However, the FWHM of C\,IV needs
to be used with caution because, not only does scattering cause
blueshifting, but it also {\em broadens} the lines (see Fig.~1). The
width of C\,IV can therefore give an overestimate of the virial
velocity in the C\,IV emitting region. Since it is the square of the
line width which enters into the equation for the virial mass,
significant systematic errors in the black hole mass can be introduced.  We have noted above
that the blueshifting increases with accretion rate, and that it has
also been found to be correlated with radio loudness and the
presence of broad absorption lines, so there is a danger of
systematic errors when using the FWHM of C\,IV to estimate masses.
It should, however, be possible to empirically correct for these
systematic errors by investigating the differences in masses
estimated from high- and low-ionization lines as a function of the
blueshifting.

\subsection{Blueshifts of Narrow Lines}

It has long been known \citep{burbidge+59} that {\em narrow} lines
in AGNs are blueshifted, and it has been widely assumed that this is
a consequence of outflow of the NLR gas and dust (e.g.,
\citealt{dahari_derobertis88}).  \citet{zamanov+02} showed that the blueshifting
of [O\,III] $\lambda$5007 is correlated with the blueshifting of broad C\,IV line
and that both blueshiftings are strongest NLS1s.  This is confirmed by
\citet{komossa+08} who have also shown that the blueshiftings of the narrow lines
increase with ionization potential. \citet{corbin90,corbin92} found that the blueshiftings of
C\,IV were related to the C\,IV Baldwin effect \citep{baldwin77,baldwin+78}.  This is also
very obvious in Fig.\@ 4 of \citet{richards+02}.  \cite{zhang+11}
show that the O\,[III] blueshiftings are similarly related to the O\,[III] Baldwin effect
The NLR blueshiftings thus show many striking similarities to the BLR
blueshiftings.  \citet{zamanov+02} argue that the correlation of blueshiftings
points to a kinematic connection between the BLR and NLR.  We therefore
suggest that if the BLR blueshiftings are due to inflow and scattering, then so too are
the NLR blueshiftings.

\subsection{Broad Absorption Line QSOs and Orientation Effects}

\citet{corbin90} discovered that his sample of BALQSOs showed unusually large blueshiftings of C\,IV
with respect to Mg\,II ($\sim 1700$ km s$^{-1}$).  Furthermore, he found that the BALQSOs
also deviated from the correlation between blueshifting and the Baldwin effect found for non-BALQSOs.
This is in the sense that their blueshifts are too large for their equivalent widths. This suggests that there is
some additional factor influencing the blueshifting.  Since BALQSOs are believed to be seen at higher inclinations \citep{hines+wills95,goodrich+miller95,cohen+95,elvis00} orientation could be the factor enhancing or even causing
the blueshifting
of high-ionization lines in BALQSOs \citep{richards+02}.  Our $STOKES$ simulations verify that in our scattering+inflow model there is indeed an enhancement of the blueshifting at high inclinations.    We find a negligible dependence of the blueshifting on viewing angle within the half-opening angle of the torus (i.e., from any type-1 viewing position), but that there is a strong increase of the blueshifting as we start seeing through the scattering medium (see Fig.~7).  The increase in blueshift is because the scattered light becomes more important.  In our models this is a relatively abrupt transition (i.e., from all type-1 viewing positions the predicted line profile is like the solid black line in Fig.~7 while for all lines of sight passing through the scattering disk the profiles are like the blue dashed line).  In a real AGN the material will be clumpy and the transition will probably not be as abrupt.

% Fig.\@ 7
\begin{figure}
\vspace*{0.3cm}
\includegraphics[width=84mm]{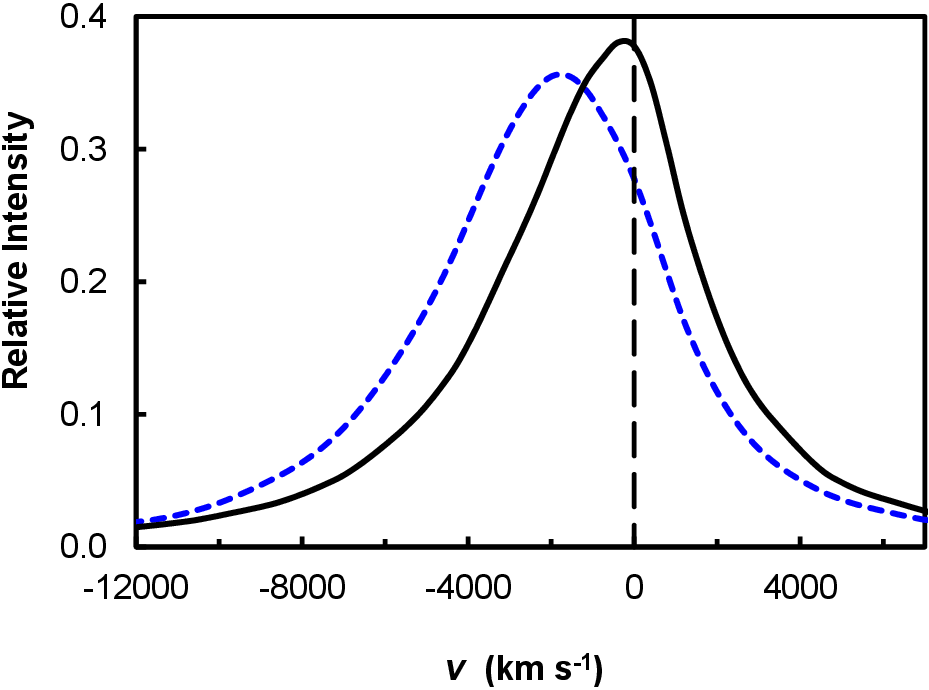} %\vspace*{0.3cm}
\caption{Theoretical C\,IV line profiles for electron or Rayleigh
scattering off a inflowing flared disk of half-opening angle 60 degrees, an inflow velocity of 1000 km s$^{-1}$, and a radial optical depth of 5. The solid black curve shows the profile
predicted when the system is viewed away from the disk. The
dashed blue line shows the profile predicted when the system is viewed from
just inside the disk surface as could be the case in a BALQSO.} \label{BALQSO}
\end{figure}
% This figure comes from line_flux_profiles.xls sheet "Plotting 2012"

For the models show in this paper we have used an unchanging unscattered profile, but since observations strongly point to the BLR being flattened and rotating (see introduction) the unscattered profile will actually be broader when seen at higher inclinations.  Our model thus also predicts that when high-ionization lines are highly blueshifted because the inclination is higher, there will be further broadening in addition to that caused by the scattering.    Inspection of Table 2 of \citet{richards+02} shows that the more blueshifted C\,IV lines are indeed broader.

Note that we do {\em not} predict that a large blueshift will be observed in type-2 AGNs.  This is because the central region is, by definition, obscured by the torus and if the BLR is detectable in polarized light, the scatterers sending light into our line of sight are located far from the central region.

\subsection{Determining the Structure of the Scattering Region(s)}

The ease with which scattering off inflowing material with differing geometries
reproduces the observed blueshifting provides gratifying support for our inwardly spiraling
BLR picture.  However, a disappointment is that
the blueshifting is insensitive to the geometry (see Figs.\ 1 and 2), and because there
is some degeneracy between the inflow rate, covering factor, and optical depth, we
cannot get a strong constraint on the geometry and kinematics
because of the limited accuracy with which line profiles can be
measured. Fortunately, other observations provide constraints on the
geometry and kinematics.  Two promising ways of studying the
distribution and kinematics of the scatters are spectropolarimetry
(e.g., \citealt{smith+04,smith+05,goosmann_gaskell07,marin+12}) and
polarimetric reverberation mapping (\citealt{gaskell+12}).  The combination of the two methods (i.e.,
high-resolution spectropolarimetric reverberation mapping) promises
to be particularly powerful.

\section{Conclusions}

We have pointed out the problems with the popular outflow/wind
explanation \citep{gaskell82} for the blueshifting and blueward asymmetry of the
high-ionization lines, and shown that velocity-resolved
reverberation mapping supports the velocity field of the BLR having an inflow
velocity component.  We have demonstrated using the {\it STOKES} Monte
Carlo radiative transfer code that electron or Rayleigh scattering off
an inflowing medium readily reproduces the blueshifts and asymmetries of the high-ionization
lines. Our model also predicts that the relative blueshifts for
different lines in the same AGN should be proportional to the
inverse square root of the radius the lines are expected to be
formed at.  Available estimates of relative blueshiftings support
this.

If the BLR is indeed inwardly spiraling this has many important implications.  Viscosity is
required to transport angular momentum outwards.  As with
traditional accretion disks, this is presumably due to the
magneto-rotational instability.  As the BLR inflows it releases
energy.  The deduced mass inflow rates are comparable to the mass
accretion rate needed to power AGNs.  We have therefore proposed
that the broad-line region could be a major part of the material accreting onto
the black hole.  Taken together these conclusions suggest a picture
where {\em the BLR is part of the outer region of the accretion disk.}

We have argued that the magnitude of the high-ionization
blueshifting effect is proportional to the mass-inflow rate. The
inflowing BLR picture thus naturally explains why high-accretion-rate AGNs (NLS1s) show the largest high-ionization-line blueshifts.  Our models
also gives an explanation of why BALQSOs can show higher blueshifts if they are seen at higher
inclinations.

The inwardly spiraling BLR picture supports use of C\,IV to measure black
hole masses in high-redshift AGNs but, because scattering broadens
lines as well as blueshifting them, caution is necessary when using
C\,IV line widths.  A systematic correction could be necessary as
a function of the blueshifting.

Similarities between NLR and BLR blueshiftings suggest that NLR
blueshiftings might also due to an inflow velocity component and scattering.

Finally, as \citet{korista_ferland98} have pointed out, Rayleigh
scattering could be important in AGNs in the ultraviolet, and we
have proposed a simple observational test for the degree to which
Rayleigh scattering influences the blueshifting.

\acknowledgments

RG is grateful to the Astronomy Department of the University of
Texas for its hospitality during his time as a Tinsley scholar where this work
was begun.  We would like to thank Philip Hardee, Liz Klimek, Julian Krolik, Bill Mathews,
Matias Montesinos, Ramesh Narayan, Daniel Proga, and
Greg Shields, for helpful discussions of various issues. This
research has been supported in part by the US National Science
Foundation through grants AST 03-07912 and AST 08-03883, by the Space Telescope
Science Institute through grant AR-09926.01, the GEMINI-CONICYT Fund of Chile through project N{\degr}32070017, FONDECYT of Chile through project N{\degr} 1120957, French grant ANR-11-JS56-013-01, the French {\it GdR} PCHE, and from the Center for Theoretical Astrophysics (CTA) through Czech Ministry of Education, Youth and Sports programme LC06014.   Finally, we would like to thank the anonymous referee for his/her helpful comments.

%\clearpage

\end{document}